\documentclass[aps,prl,twocolumn,groupedaddress,showpacs,floatfix]{revtex4}

\usepackage{latexsym,amsmath,amscd,amssymb,graphics}
\usepackage{enumerate}

\usepackage{graphicx}
\usepackage{framed}
\usepackage[colorlinks]{hyperref}
\numberwithin{theorem}{section}

\def\be{\begin{equation}}
\def\ee{\end{equation}}
\def\bea{\begin{eqnarray}}
\def\eea{\end{eqnarray}}
\def\ba{\begin{array}}
\def\ea{\end{array}}

\def\bOm{\boldsymbol{\Omega}}

\let\<\langle
\let \>\rangle
\newcommand{\rem}[1]{}

\newcommand{\bq}{\boldsymbol{q}}

\newcommand{\br}{\boldsymbol{r}}
\newcommand{\bGam}{\boldsymbol{\Gamma}}
\newcommand{\bom}{\boldsymbol{\omega}}
\newcommand{\bgam}{\boldsymbol{\gamma}}

\newcommand{\bkappa}{\boldsymbol{\kappa}}

\newcommand{\blam}{\boldsymbol{\lambda}}

\newcommand{\pp}[2]{\frac{\partial #1}{\partial #2}}

\newcommand{\lp}{\left(}
\newcommand{\rp}{\right)}

\begin{document}

%\def\below#1#2{\mathrel{\mathop{#1}\limits_{#2}}}

%%%%%%%%%%%%%%%%%%%%%%%%%%%%%%%%%%%%%%%%%%%%%%%%%%%%%%%%%%%%%%%%%%%%%%%%%%%%%%%
%%%%%%%%

%%%%%%%%%%%%%%%%%%%%%%%%%%%%%%%%%%%%%%%%%%%%%%%%%%%%%%%%%%%%%%%%%%%%%%%%%%%%%%%
%%%%%%%%

\title{Dynamics of elastic rods in perfect friction contact}
\author{Fran\c{c}ois Gay-Balmaz$^1$ and 
Vakhtang Putkaradze$^2$ \\ 
$^1$ LMD - Ecole Normale Sup\'erieure de Paris - CNRS, 75005, Paris\\ 
$^2$ Department of Mathematical and Statistical Sciences \\ 
University of Alberta, Edmonton, AB   T6G 2G1 Canada
}

\date{ \today }
\pacs{46.70.Hg, 46.40.-f, 87.85.gp, 87.15.ad}
%\makeatother

\begin{abstract}
One of the most challenging and basic problems in elastic rod dynamics is a
description of rods in contact that prevents any unphysical self-intersections. Most
previous works addressed this issue through the introduction of short-range
potentials. We study the dynamics of elastic rods with perfect rolling contact which
is physically relevant for rods with rough surface, and cannot be
described by any kind of potential. We derive the equations of motion and show that the
system is essentially non-linear due to the moving contact position, resulting in a
surprisingly complex behavior of the system.
\end{abstract}

\maketitle

\begin{figure} 
\includegraphics[width=6cm]{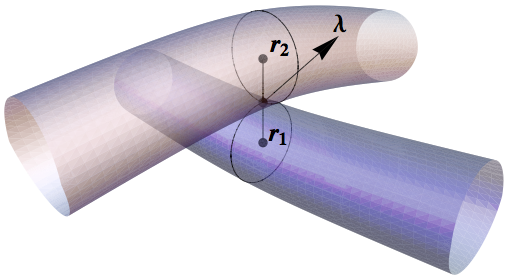}
\caption{A sketch of two strings in contact. The centers of disks in contact are marked $\br_1$ and $\br_2$, the line connecting them passes through the contact point. Reaction force of the contact is also shown as $\blam$.  } 
\label{fig:contact}
\end{figure}

\noindent 
{\bf Introduction} 
Take two rubber strings, stretch them a bit and cross them so they remain in contact, as shown on Figure~\ref{fig:contact}.  As long as the deformations are not too large, the strings will roll at the contact without sliding. Of course, that simple experiment  is dominated by the energy loss from the internal deformation of strings at the contact point; however, improving the quality of  strings will allow them to oscillate for a reasonable time before the energy loss takes over. Interestingly,  this simple and familiar experiment has deep mathematical and physical implications that go beyond a toy problem. 

There are many objects that can be represented as long elastic rods, from a rubber hose to DNA molecules. Typically, if these objects are put in a confined space, or undergo other non-trivial dynamics, self-contact of these rods typically appears. The true dynamics will not allow the rod to pass through itself, and it must preserve the side of contact under dynamics. 
 While it is generally accepted that something like DNA at contact will slide freely, the dynamics of other molecules like dendronized polymers (DP) may be different.  These compound molecular structures are formed by assembling multiple \emph{dendrons} that are each connected by its base to a long polymeric backbone  \cite{ToNaGo1990}. A simplified, coarse-grained rod model of such polymers must take into account the rough surface formed by tree-like structures that is likely to generate perfect rolling contact. 

Efficient numerical methods have been developed recently to  deal with the self-contact forces of rods using short-range repulsive potential for statics \cite{Ba-etal-2007} and dynamics 
 \cite{GoPeLe2005,GoPeLe2008,LiPe2011}.   Another avenue of studies of \emph{stationary states} of elastic rods with self-contact   \cite{WeToOl1997,CoSwTo2000,CoSw2000,Go-etal-2002,CoSw2004} explicitly computes   the contact forces from the existence of constraints. In our case, the rolling contact comes from friction, which does not admit any potential description.  The extension of the latter approach  to include the dynamics, and especially the rolling constraint is difficult.  We also note that the true motion may be a combination of rolling and sliding friction, but in the absence of a consistent theory for rolling slippage we shall concentrate on the perfect rolling only.  Our solution of the problem of rolling contact is similar in spirit of the second approach, as the contact force comes naturally from the constraint and not from short-range forces.

\noindent 
{\bf Setup of the problem} 
As is well known \cite{Bloch2003}, the problem of rolling motion is essentially non-holonomic and, in general, cannot be represented from the potential point of view.  We shall note that it is possible to define potential approach to some non-holonomic  systems \cite{BlRo2008}. 
However, these cases seem to be more of an exception rather than a rule. For the problem in hand, we proceed with the Lagrange-d'Alembert (LdA) principle which is the fundamental method for the treatment of non-holonomic systems \cite{Bloch2003,LeMu1994}. Similar approach  has been recently used to describe the motion of an elastic rod rolling on a plane \cite{Va2007}.  
In order to utilize LdA, we recast the motion of the rods in variational setting through the Simo-Marsden-Krishnaprasad (SMK) rod theory \cite{SiMaKr1988}, which allows a variational formulation of string dynamics \cite{ElGBHoPuRa2010}.  The rod is parameterized by a coordinate $s$ that does not have to be arc length. We fix a reference frame and measure the local position $\br(s,t)$ and orientation $\Lambda(s,t)$ at time $t$ with respect to the fixed frame. SMK theory formulates the dynamics in terms of the variables that do not depend on the choice of basic frame:  
\begin{align}
\bGam=\Lambda^{-1} \br' \, , \, 
\bOm=\Lambda^{-1} \Lambda' \, , \,  
%\label{invar1} 
%\\ 
\bgam=\Lambda^{-1} \dot\br \, , \, 
\bom=\Lambda^{-1} \dot \Lambda \, . 
\label{invar}
\end{align} 
Here, the prime denotes the partial derivative with respect to $s$ and dot the derivative with respect to $t$. The physical meaning of $\bgam$ and $\bom$ is the linear and angular velocity in the body frame of reference, and $\bGam$ and $\bOm$ the corresponding deformation rate. If $\ell(\bGam, \bOm,\bgam,\bom)$ is the Lagrangian, then the equations of motion  for a free elastic rod are given by the variational principle \cite{ElGBHoPuRa2010} 
\vspace{-2mm}
\begin{equation} 
\delta \int \ell(\bGam, \bOm,\bgam,\bom) \mbox{d} s \mbox{d} t=0 \, , \quad 
\label{deltaS}
\vspace{-2mm}
\end{equation} 
 appropriately computing the variations of variables $(\bGam, \bOm,\bgam,\bom)$. 
For purely elastic rods, $\ell$ is a quadratic function; more complicated expressions for $\ell$ are also possible. 
In the case of constraints, this variational principle (\ref{deltaS}) has to be modified according to the LdA approach as follows. Assume, for simplicity, that the undeformed rod has a circular cross section and the acting forces  are small enough so that the cross-section remains circular even at contact.  Thus, if the strands at contact can be approximated by touching circular cylinders, then the contact point is always located at 
$c(t)=(\br_1+\br_2)/2$, and the vector from the center of the cylinder $\br_i(t) $ to the contact point is 
$\pm (\br_2-\br_1)/2$. 
Here, the index $i=1,2$ means evaluation at $s=s _i $ marking the disks at contact.   Since the angular velocity in the fixed frame is $ \dot \Lambda_i\Lambda_i^{-1}$, $i=1,2$, then the velocity of the material point associated with contact, also in the fixed frame, is
\vspace{-2mm} 
 \begin{equation}
 \dot \br_1 +
 \frac{1}{2} \dot \Lambda_1 \Lambda_1^{-1}  \big( \br_2-\br_1 \big) 
 =\dot\br_2-
  \frac{1}{2} \dot \Lambda_2 \Lambda_2^{-1} \big( \br_2-\br_1 \big) 
  \label{contactvel}
  \vspace{-2mm} 
 \end{equation}
 Note that it is not the same velocity as the velocity of the contact point, which will include the full derivatives, and, as a consequence, $\dot s_i$. 
 This condition can be reformulated in terms of invariant variables by multiplying  \eqref{contactvel} by  $\Lambda_{1}^{-1}$: 
 \vspace{-2mm} 
\begin{equation}
 \bgam_1 + \frac{1}{2} \bom_1 \times \bkappa_{12} 
= 
  \xi_{12} \bgam_2 - \frac{1}{2} \big( \xi_{12} \bom_2\big)  \times \bkappa_{12}  \, , 
\label{rollingcond}
\vspace{-2mm} 
\end{equation}
where $\xi_{12}:=\Lambda_1^{-1} \Lambda_2$ is the relative orientation, $\bkappa_{12}=\Lambda_1^{-1} (\br_2-\br_1)$, 
and other invariant variables are defined as in (\ref{rollingcond}), with the index $i$ meaning evaluation at $s=s_i$. Note that due to the uniformity of the strand, and the assumption of circular cross-section, the Lagrangian does not depend explicitly on $s_i$ and $\dot s_i$. 
Then, LdA principle states that one has to replace \emph{time derivatives only} in (\ref{rollingcond}) by $ \delta $-variations and use that expression as an additional constraint on  $(\delta \bGam, \delta \bOm,\delta \bgam,\delta \bom)$ in  \eqref{deltaS}. For example,
\vspace{-2mm} 
\begin{equation} 
\bgam_1=\Lambda_1^{-1} \dot \br_1 \rightarrow \Lambda_1^{-1} \delta \br(s)
 \delta_{s_1} \, , 
 \label{deltagam1}  
\vspace{-2mm} 
\end{equation}
and similarly for other variables. Here we have denoted, for shortness, $\delta_{s_i}:=\delta(s-s_i)$. There is an unfortunate collision of notation in $\delta$ between the variational derivatives and Dirac's $\delta$-functions; in this paper, the $\delta$-function always has a subscript. Note that one does not enforce the condition \eqref{rollingcond} as it stands: this approach leads to the so-called \emph{vakonomic} approach \cite{ArKoNe1997} which can be shown to give incorrect equations of motion for this case.   

Denote by $\blam(t)$ a vector that enforces the LdA constraints, and  by  ${\rm D}/{\rm D} t:=\partial_t + \bom \times $, ${\rm D}/{\rm D} s:=\partial_s+ \bOm \times$ the full $t$- and $s$- derivatives. 
We get the following equations for the motion of  strings with rolling contact: 
\vspace{-1mm} 
\begin{align}
   \frac{{\rm D}}{{\rm D} t}  \frac{\partial  \ell }{\partial  \bgam} &
+
\frac{{\rm D}}{{\rm D} s } \frac{\partial     \ell }{\partial  \bGam}
= 
\blam \delta_{s_1}- \xi_{12}^{-1} \blam \delta_{s_2} 
\label{EPpsi}
\\
 \frac{{\rm D}}{{\rm D} t}   \frac{\partial  \ell}{\partial  \bom} &
+
\frac{{\rm D}}{{\rm D} s} \frac{\delta \ell}{\delta \bOm}
=  \frac{\partial  \ell}{\partial  \bgam} \times \bgam
+
\frac{\partial   \ell   }{\partial  \bGam} \times \bGam
\nonumber 
\\ 
& 
+ \frac{1}{2} \bkappa_{12}\times  \blam \delta_{s_1}
+\frac{1}{2} \xi_{12}^{-1} \Big( \bkappa_{12} \times  \blam   \Big)  \delta_{s_2} .
\label{EPsigma}
\vspace{-2mm}
\end{align}
The physical meaning of  $\blam$ as the force due to the constraint is now evident from the linear momentum \eqref{EPpsi}.  Correspondingly, the $\blam$ terms in equation \eqref{EPsigma} are identified as torques acting on the contact point due to the presence of the constraint.  These equations have to be augmented by the compatibility conditions 
\vspace{-2mm} 
\begin{equation} 
\begin{aligned} 
&\dot \bGam = \bgam'-\bom \times \bGam+\bOm \times \bgam \\ 
&\dot \bOm = \bom'+ \bom \times \bOm \, . 
\end{aligned} 
\label{compatibility}
\vspace{-2mm}  
\end{equation} 
We still have to close the system by computing the equations of motion for the contact points $s_i$, which is done from the tangency conditions stating that the strand at a contact point, which is locally a cylinder, touches itself tangentially and there is no intersection of the material. These conditions state that 
\vspace{-1mm} 
\begin{equation} 
\begin{aligned} 
&\bkappa_{12} (s_1,s_2,t) \cdot \bGam_1(s_1,t)=0, 
\\
&\bkappa_{12}  (s_1, s_2,t) \cdot \xi_{12} \bGam_2 (s_2,t)=0.
\end{aligned}  
\label{tangency} 
\vspace{-1mm} 
\end{equation}
Note that these expressions do not contain any time derivatives of the variables and are thus \emph{holonomic}.  They also imply that the distance between the centers of  disks in contact $|\br_1-\br_2|$ is preserved.  Since the Lagrangian $\ell$ does not depend on $s_i$ and $\dot s_i$,  they can be imposed after the equations have been derived.  In principle, such conditions already determine $s_i$ through an implicit relation; however, such relation is difficult to use. Instead, we take the time derivatives of the tangency conditions to obtain a relationship  
\vspace{-2mm} 
\begin{equation} 
\mathbb{A} \cdot (\dot s_1 \, , \, \dot s_2)^T + \mathbf{v}=0\, , 
\label{s1s2dot} 
\vspace{-2mm} 
\end{equation} 
where 
the 2$\times$2 matrix $\mathbb{A}$ and 2-vector $\mathbf{v}$ depends on the dynamical properties at contact, with ${\rm det} (\mathbb{A}) \neq 0$ when the rods are not locally parallel at the contact. 
\rem{ %%%BEGIN REM 
\begin{widetext} 
\begin{align} 
0&=\pp{\bkappa_{12}}{t} \cdot \bGam_1 + \bkappa_{12} \cdot  \pp{\bGam}{t}
+
\Big( 
\pp{\bkappa_{12}}{s_1} \cdot \bGam_1 + 
\bkappa_{12} \cdot \pp{\bGam_1}{s_1}
\Big) \dot s_1 
+ 
\pp{\bkappa_{12}}{s_2} \cdot \bGam_1 \dot s_2  
\label{s1doteq} 
\\
0&=\pp{\bkappa_{12}}{t} \cdot \xi_{12} \bGam_2 + 
\bkappa_{12} \cdot \pp{ \xi_{12}  }{t} \bGam_2 +
\bkappa_{12} \cdot \xi_{12} \pp{  \bGam_2 }{t} +
\Big( 
\pp{\bkappa_{12}}{s_1} \cdot  \xi_{12}  \bGam_2 + 
\bkappa_{12} \cdot \pp{\xi_{12} }{s_1}\bGam_2
\Big) \dot s_1 
\nonumber 
\\ 
& \qquad \qquad 
+ 
\Big( 
\pp{\bkappa_{12}}{s_2} \cdot  \xi_{12}  \bGam_2 + 
\bkappa_{12} \cdot \xi_{12} \pp{\bGam_2}{s_2}+
\bkappa_{12} \cdot \pp{\xi_{12} }{s_2}\bGam_2
\Big) \dot s_2
\label{s2doteq} 
\end{align} 
\end{widetext} 
} %%%END REM 

\noindent 
{\bf Discrete  strands in contact} 
It is also interesting to consider the application of this theory to discrete, chain-like elastic structures in contact. In that case, we need to clarify the physical meaning of the $\delta$-function at the contact position.  Apart from its physical relevance, this consideration is also useful for consistent numerical discretization of  (\ref{EPpsi},\ref{EPsigma}).  Here, care must be taken in deriving the equations of motion without breaking their variational  structure  \cite{MoVe1991}. 
Suppose that we have a string consisting of discrete set of points along the line, $s=s^k$, with $k$ being integer.  If the orientation and position of a material frame at $s=s^k$ are given by an orientation matrix $\Lambda_k$ and a vector $\br_k$,  the invariant variables are 
$
p_k=\Lambda_k^{-1} \Lambda_{k+1} 
$ and 
 $\bq_k = \Lambda_k^{-1} (\br_{k+1}-\br_k) 
$.
The purely elastic Lagrangian is $\ell= \ell(\bom_k, \bgam_k, p_k, \bq_k)$. One also needs to define a "smeared-out" version of \eqref{rollingcond}:
\vspace{-2mm}
\begin{align}
  \alpha_k   \big( &\bgam_k +\frac{1}{2}   \beta_m \bom_k \times \bkappa_{km} \big)  
  \nonumber 
  \\ 
 - & \alpha_k \beta _m \big(  \xi_{km} 
  \bgam_m -\frac{1}{2}  (\xi_{km} \bom_m) \times \bkappa_{km} \big) =0 \, , 
\label{avrolling2} 
\vspace{-6mm}
\end{align} 
(summation over $k,m$). Here, 
 we have defined the averaging coefficients: $\alpha_k:=G\big(s_1-s^k\big)$ $\beta_k:=G\big(s_2-s^k\big)$ arising from a "bump" function $G(s)$ that rapidly decays away from $s=0$, and 
$ \xi_{km}:=\Lambda_{k}^{-1} \Lambda_m$,   
$\bkappa_{km}:=\Lambda_k^{-1} \big( \br_m-\br_k \big)$.  
 Note that the positions of the contact is defined by the continuous variables $s_1$ and $s_2$. 
The physical meaning of \eqref{avrolling2} consists in spreading the point wise contact condition \eqref{rollingcond} to a few neighboring points surrounding the contact. Then, the LdA principle gives a discrete analogue of (\ref{EPpsi},\ref{EPsigma}): 
\begin{align}
& \! \frac{ {\rm D} }{ {\rm D} t}  \frac{\partial \ell}{\partial \bgam_k}
\!-\!
p_{k-1}^{-1} \!  \frac{\partial \ell}{\partial \bq_{k-1} \!} 
\! + \!
\frac{\partial \ell}{\partial \bq_k}
= \! \! 
\sum_m 
\! \!
 \big( \alpha _m  \beta _k  \xi_{mk}^{-1}\! -\! \alpha_k \big)  \blam , 
 \label{EPsigma-discr}
\\
&\frac{ {\rm D} }{ {\rm D} t} \frac{\partial  \ell }{\partial  \bom_k}  
+
\frac{\partial  \ell}{\partial p_k} p_k^{-1} -
p_{k-1}^{-1} \frac{\partial  \ell}{\partial p_{k-1}} 
=\! 
 \frac{\partial \ell }{\partial \bgam_k}\!  \times\!   \bgam_k
\! + \! 
\frac{\partial l}{\partial \bq_k} \! \times\!   \bq_k 
\nonumber 
\\
&\quad 
\!-\!
\sum_m 
\frac{1}{2} \left(  \alpha_k \beta_m \bkappa_{km} \! \times \! \blam 
\!+ \! \beta_k  \alpha_m  
 \xi_{mk}^{-1} \big( 
 \bkappa_{mk} 
\! \times\! \blam\big)\right) .
 \label{EPpsi-discr}
 \vspace{-2mm}
\end{align} 
It is interesting to note that for a Lagrangian that is quadratic in all variables (\emph{e.g.}, linear elasticity), (\ref{EPsigma-discr},\ref{EPpsi-discr}) allow explicit calculation of the constraint $\blam$. 
These equations have to be augmented by conditions for the variables $s_1$ and $s_2$. The discrete version of \eqref{tangency} is  
$ 
\sum_{lm} \alpha_l \beta_m \bkappa_{lm} \cdot  \bGam_l  =0
$ and 
$\sum_{lm} \alpha_l \beta_m \bkappa_{lm} \cdot   \xi_{lm} \bGam_m =0
$. 
Differentiating  these expressions with respect to $t$, we again obtain the equations governing the evolution of derivatives $\dot s_1$ and $\dot s_2$ as 
$\mathbb{A}_d \cdot (\dot s_1 \, , \, \dot s_2)^T + \mathbf{v}_d=0$, where the matrix $\mathbb{A}_d$ and the vector $\mathbf{v}_d$ depend on the dynamical variables close to the contact.

\noindent 
{\bf Linear strings in contact} 
One may wonder if the equations of motion we have derived allow to deduce analytical expressions for the propagation of the disturbances along the rods at contact.  The answer to this question is, unfortunately, no: the contact condition makes the disturbances essentially nonlinear. 
Let us consider two strings in contact, and denote for shortness $\mathbf{a}= (\bgam,\bom)^T$  
and $\mathbf{A}= (\bGam,\bOm)^T$. For linear elastic materials 
$ \partial \ell/\partial \mathbf{a} = \mathbb{V} \mathbf{a}$ and 
$ \partial  \ell /\partial \mathbf{A} = - \mathbb{Q} \mathbf{A}$, 
where $\mathbb{V}$ and $\mathbb{Q}$ are 6x6 matrices.  The linearized compatibility conditions \eqref{compatibility} allow to introduce a vector potential $\boldsymbol{\phi}$ as 
$\mathbf{a}=\partial_t \boldsymbol{\phi}$, $\mathbf{A}=\partial_s \boldsymbol{\phi}$. 
Neglecting all nonlinear terms in the dynamic variables  and  assuming that the rod is naturally straight in its undeformed state, we can transform  (\ref{EPpsi},\ref{EPsigma}) 
into a vector  wave equation  \cite{BiCoZh2004}
\vspace{-2mm}
\begin{equation} 
\mathbb{V}  \frac{\partial^2 \boldsymbol{\phi} }{\partial t^2\!}
\!- \! \mathbb{Q} \frac{\partial^2 \boldsymbol{\phi} }{\partial s^2 \!} 
\!=\!
\lp
\! \! 
\begin{array}{c}
{\rm Id} \\ 
\frac{1}{2} \bkappa_{12} \!\times\!
\end{array}
\! \!
\rp 
\blam \delta_{s_1}
\! \!+\!
\xi_{12}^{-1} 
\! \! 
\lp 
\! \! 
\begin{array}{c}
-{\rm Id} \\ 
\frac{1}{2} \bkappa_{12}\! \times \! 
\end{array}
\! \! 
\rp 
\blam \delta_{s_2}. 
\label{wavephi} 
\end{equation} 
The condition \eqref{rollingcond} can be expressed in this vector form as 
\vspace{-4mm} 
\begin{equation} 
\big( {\rm Id}, \! -\frac{1}{2} \bkappa_{12} \times   \big)^T
\! 
\boldsymbol{\phi}_t (s_1,t) 
\!= \!
\big( \xi_{12} , \! \frac{1}{2} \bkappa_{12} \times \xi_{12} \big)^T\! \! \boldsymbol{\phi}_t (s_2,t) 
\label{contactcondlin} 
\vspace{-2mm} 
\end{equation} 
Thus, the evolution of small disturbances on the rod is governed by linear equations (\ref{wavephi}), linear rolling constraint \eqref{contactcondlin} and \emph{nonlinear} 
evolution equations for $s_{1,2}$ \eqref{s1s2dot}. We can illustrate the complexity of this problem on a pedagogical simplistic example of two straight rods in contact with only one mode being relevant in \eqref{wavephi} for each rod.  Let us call these modes $u(x,t)$ for the first rod and $v(y,t)$ for the second rod, respectively. Such a simple model illustrates the full  complexity of the problem without the unnecessary complications of higher dimensionality of $\boldsymbol{\phi}$. Physically,  such a mode can be realized for a rod with special elastic and inertia matrices $\mathbb{V}$ and $\mathbb{Q}$, \emph{e.g.} for rods made out of composite material.  In this case, the rolling constraint and the motion of the contact points  are  written simply as 
\vspace{-2mm}
\begin{equation} 
\label{constr0}
u_t(s_1) =F v_t(s_2) \, , \quad  \dot s_i = G_i u_t (s_i) \, , \, i=1,2 \, , 
\vspace{-2mm}
\end{equation} 
where $F$  and $G_i$ are constants depending on the material parameters and the base state of the rods. 
The equations of motion for $(u,v)$ in this reduced setting are: 
\vspace{-2mm}
\begin{equation} 
u_{tt}-c^2 u_{xx}+\lambda \delta_{s_1}  =0\, , \,  % \label{ueq}
v_{tt}-c^2 v_{yy}-\lambda F \delta_{s_2}  =0, \label{uveq}
\vspace{-2mm}
\end{equation} 
where $\lambda$ enforces the first constraint of \eqref{constr0}. 

A discrete version of these equations can be derived, similarly to the full equation described above in (\ref{EPpsi-discr}). 
We shall emphasize that the complexity caused by the nonlinear contact conditions is the same in the full equation \eqref{wavephi} and its one-dimensional counterpart (\ref{uveq}). 
 While the complex dynamics caused by the nonlinear rod equations have been well studied, as far as we are aware, there has been no work on the complex dynamics caused by the rolling contact condition. In the absence of constraints, the wave equations provide a simple harmonic oscillation of the string. However, when the constraint is present,  the motion of the string is challenging and complex. 

\noindent
{\bf Wave equation: exact solution and contact chaos} 
In the case when $u(x,t)$ and $v(y,t)$ have periodic boundary conditions on an interval of the same length, an exact solution of the equation  (\ref{uveq}) can be found, in spite of the contact nonlinearity. 
Let us consider the case when the strings are identical, so $G_1=G_2=G$ and $F=1$.  In that particular case, from \eqref{constr0}, we conclude that $s_2 - s_1=d=$const. By shifting the  coordinate $y$ in the $v$-equation of \eqref{uveq} as $y=x+d$, using periodic boundary conditions and denoting the new coordinate $x$ so both $u$ and $v$ depend on  the same variables,  we can  choose $s_1(0)=s_2(0)$. From $\dot s_1 = \dot s_2$,  we get $s_1(t)=s_2(t)=s(t)$.  
Define 
$
P =u+v
$
and 
$
M=u-v
$,   satisfying
\vspace{-1mm}
\begin{align} 
&P_{tt}-c^2 P_{xx} = 0\, , \, 
M_{tt}-c^2 M_{xx} = 2 \lambda  \delta \big(x-q (t) \big) 
\label{PMeq} 
\\ 
&M_t (S) = 0 \, , \qquad \dot s = \frac{1}{2} G P_t (s)  \, . 
\label{dotqeqP} 
\vspace{-3mm}
\end{align}
As we see, equation \eqref{PMeq} for $P(x,t)$  is now decoupled and can be solved independently from the other equations as a simple wave equation on the circle. Then, $s(t)$ is obtained by solving the first-order differential equation \eqref{dotqeqP}. To find 
the solutions $u(x,t)$ and $v(x,t)$, one has to solve equation \eqref{PMeq} with prescribed  $s(t)$ which is a much simpler problem. 
The solution $P(x,t)$ can be any function satisfying a wave equation with periodic boundary conditions. 
The equation for $s(t)$ is then 
$\dot s = G P(s(t), t)/2$ which is a nonlinear, non-autonomous equation; it is well known that such equations lead to rather complex behavior for $s(t)$ even if $P(x,t)$ is a simple function. However, this behavior is not truly chaotic in our opinion. 

\begin{figure}[h]
\includegraphics[height=4.2cm]{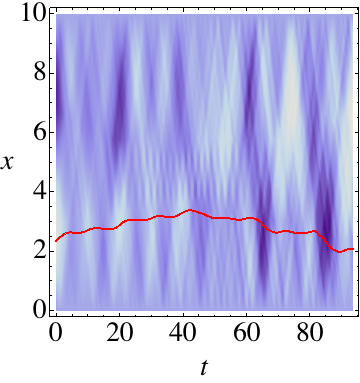}
\includegraphics[height=4.2cm]{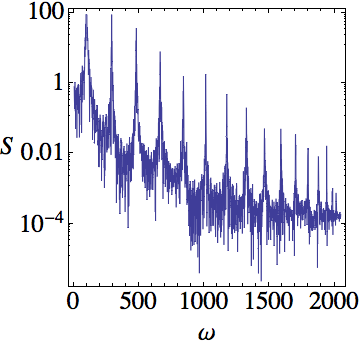}
\caption{Left: the spatio-temporal evolution of one of the strings in contact, with the red line marking the point of contact.  Right:  spectrum $S(\omega)$ of time signal produced by the lead harmonic (in $x$) of $u(x,t)$.  } 
\label{fig:violin}
\end{figure}

The situation is much more interesting for the case of more realistic  fixed boundary conditions for $u$ and $v$.  This would be the case of two rods with fixed ends in contact.  Such motion, as one can show from equations (\ref{uveq}), conserves energy; however, in reality, friction with air and, more importantly,  rolling friction will lead to energy dissipation. It is nevertheless interesting to see the structure of that dynamics,  with the nonlinearity obtained only from the contact condition. As we see from Figure~\ref{fig:violin}, the system produces a complex spatio-temporal dynamics of both strings. It is also relevant to present another measure of complexity, computed from the dynamics of the base harmonic of $u(x,t)$, call it $\hat{u}_1(t)$,  as a function of $t$. If the string were vibrating in the air, the sound sufficiently far away from the string will primarily contain the contribution from the first harmonic. 
On Figure~\ref{fig:violin}, right, we plot the time spectrum $S(\omega)$ as a function of temporal frequent $\omega$, 
 obtained from the time signal of the first harmonic $\hat{u}_1(t)$.
 Starting with $u(x,0)=\sin x$, a linear rod with $0<x<2 \pi$ without rolling contact will generate a purely monochromatic sound; however, when the contact is present, there is a persistence of high overtones to the signal. We have found out that the chaotic behavior caused by the contact condition persists for all initial conditions we have tried. We believe that the appearance of the chaos due to the contact is highly interesting and important for many physical applications, and yet it has not been discussed in previous literature. 

\noindent 
{\bf Acknowledgements} 
We have benefitted from inspiring discussions with D. D. Holm, T. Ratiu and C. Tronci. 
FGB was partially supported by a ``Projet Incitatif de Recherche" contract from the Ecole Normale Sup\'erieure de Paris and by the Swiss NSF grant 200020-137704.
This project received support from the Defense Threat Reduction Agency -- Joint Science and
 Technology Office for Chemical and Biological Defense (Grant no. HDTRA1-10-1-007).  
 VP was partially supported by the grant NSF-DMS-0908755, the University of Alberta Centennial Fund. This project has also received support from the WestGrid at the University of Alberta. 
\vspace{-3mm}

{\footnotesize 
\bibliographystyle{unsrt}
\bibliography{papers-strands}
}
\end{document}